# DiME and AGVis: A Distributed Messaging Environment and Geographical Visualizer for Large-scale Power System Simulation

Nicholas Parsly[1], Jinning Wang[1], *Student Member, IEEE*, Nick West[1], Qiwei Zhang[1], *Member, IEEE,*
Hantao Cui[2], *Senior Member, IEEE*, Fangxing Li[1], *Fellow, IEEE*

[1]Department of Electrical Engineering and Computer Science, University of Tennessee, Knoxville, Knoxville, TN

[2]School of ECE, Oklahoma State University, Stillwater, OK, USA

nparsly@vols.utk.edu, jwang175@vols.utk.edu, nwest13@vols.utk.edu, qzhang41@vols.utk.edu, h.cui@okstate.edu, fli6@utk.edu

*Abstract*—This paper introduces the messaging environment and the geographical visualization tools of the CURENT Large-scale Testbed (LTB) that can be used for large-scale power system closed-loop simulation. First, Distributed Messaging Environment (DiME) is presented as an asynchronous shared workspace to enable high-concurrent data exchange. Second, Another Grid Visualizer (AGVis) is presented as a geovisualization tool that facilitates the visualization of real-time power system simulation. Third, case studies show the use of DiME and AGVis in power system research. The results demonstrate that, with the modular structure of DiME and AGVis, the LTB is capable of not only federal use for real-time, large-scale power system simulation, but also independent use for customized power system research.

## I. Introduction

CURENT LTB [1] facilitates power system prototyping and simulation as a large-scale digital twin. With a modular structure, the LTB consists of a series of independent open-source packages, including ANDES, DiME, and AGVis. LTB features large-scale simulation, communication, and geographical visualization. An open-source power system dynamic simulator, ANDES using a hybrid symbolic-numeric framework [2]-[3], is developed for rapid prototyping. Besides the simulation engine ANDES, a messaging tool is needed for other power grid modules such as the geographical visualization and the energy management system.

Recently, the libraries for power system studies have been enriched by the research community, from steady state analysis [4]-[6] to dynamic simulation [7]-[9]. These libraries can be employed in a distributed manner to alleviate the computation and communication burden when scaling up for a large-scale system. Open MPI [10] is an available open-source messaging package that can be used for distributed messaging; however, it can be bottlenecked by massive data volumes and heterogeneous programming languages. Geographical visualization, also known as geovisualization, is widely applied in fields like climatology [11] and archeology [12] because of its intuitive and understandable display of information. While there exists some popular geovisualization software, either commercial or open-source, such as ArcGIS [13] and Google Earth [14], they tend to fall short for visualizing a real-time, large-scale power system simulation. This is due to the fact that the challenges in power system simulations are usually specific to a given power system, such as performance with massive data and interfacing with other power system simulation engines.

While existing tools can be used for prototyping large-scale power systems, few were made with that express purpose and mind, and thus fail to fit the specific needs of researchers. To address the challenges of data exchange and visualization for real-time power system simulations or digital twins, this paper introduces two open-source packages, DiME and AGVis, which are packages within CURENT LTB as well as independent modules. The main contribution of this paper is summarized as follows:

● DiME has been developed to enable high-concurrency, high-volume real-time data exchange in large-scale power system simulations using a shared workspace. Furthermore, DiME is compatible with multiple programming languages, which allows the application of other power system simulation tools.

● AGVis, a geographical visualizer with a "MultiLayer" feature, has been developed to facilitate geographical interpretation of power system simulation results. Further, the enhanced AGVis allows not only for real-time visualization for data coming from ANDES, but also for independent use with user-defined data.

● This work demonstrates not only the federal use of ANDES, DiME and AGVis for real-time, closed-loop virtual power grid simulation at a large-scale, but also the use of AGVis for user-provided data and multi-energy systems. These demonstrations show how AGVis and DiME can both quickly and conveniently provide easily interpretable visualizations for interpreting power systems beyond just the data output by a simulation.

The rest of the paper is organized as follows: Section II discusses the philosophy of the asynchronous shared workspace that is implemented in DiME. Section III explains the geographical and MultiLayer visualization of AGVis. Section IV presents case studies for DiME and AGVis. Section V concludes this paper.

## II. DiME

The version of DiME that has been added to LTB is the second generation of DiME. The original DiME server was written in Python, and there were only two DiME clients. Although this original version performed quite admirably, it ultimately ran too slowly when working with larger simulations. Thus, DiME was fully rewritten to improve its speed, portability, and usability.

### A. DiME Server and Clients

Generally, traditional programming languages run substantially faster than scripting languages. This is one of the primary reasons why DiME was rewritten in C as opposed to its original Python, the other being to reduce dependencies. Depending on how it is initiated, a DiME server can handle connections using IPC, TCP, and WebSockets. DiME uses a simple *poll(2)* [15] loop to handle network connections. Then, depending on a user's operating system, it either uses the *libev* [16] library or the *WinSock2* [17] library to handle synchronous I/O events. Despite the simplicity of DiME's I/O handling, it has throughput of 26 million IEEE doubles per second. This is a speedup of approximately forty times compared to the first generation of DiME, which could only handle about 650,000 IEEE doubles per second. Although additional libraries or multithreading could admittedly be used to improve DiME's throughput further, the simplicity of DiME's current system speaks for itself and fulfills the needs of CURENT's simulations.

The current release of DiME is compatible with three programming language clients—MATLAB, Python, and JavaScript. These are all made to share approximately the same functionality and are compatible with the same variable types. They can not only send and receive standard data types like strings, integers, and doubles, but also handle complex numbers and NumPy-style N-dimensional arrays [18]. The only major limitation any of clients is that the JavaScript client can only use WebSockets. Other than that, though, they all effectively have the same features.

### B. Shared Workspace

The shared workspace distributed computing model that DiME implements allows clients to share variables between other clients using a table maintained by the server. The server keeps track of several tables owned by different groups. When a client connects to a DiME server, it can request a group to join by name. If this group is not available, the server will create it and its associated table. Clients can join more than one group at a time. A client can update the server's tables by passing variable names to either the *send()* function, which can specify which groups' tables to update, or the *broadcast()* function, which updates the passed variables for all tables on the server. The *send_r()* and *broadcast_r()* functions can also send data, but require the user to specify what variable they want to send and provide a value to send. Clients are able to receive data by calling the *sync()* method, which will update client-side variables, or *sync_r()*, which returns a table containing the variables and their updated values. Clients can only receive data that was passed to groups that they are a part of. Fig. 1 and Fig. 2 demonstrate the relationship between the clients, servers, and groups.

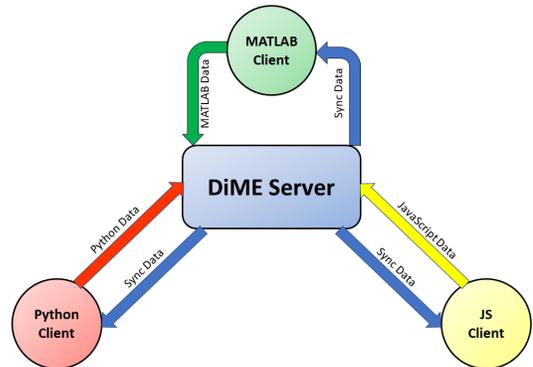

Figure 1. A representation of the interactions between DiME's server and clients.

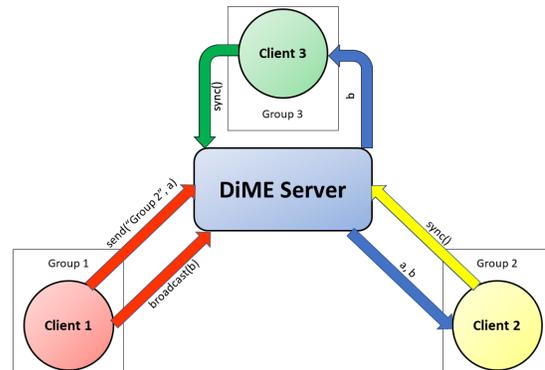

Figure 2. A representation of the difference between the *broadcast()* and *send()* commands in DiME.

## III. AGVis

AGVis is a Web-based JavaScript visualization application that runs in users' web browsers. The original intent of AGVis was to receive data through DiME and act as a visualizer for ANDES within the framework of CURENT LTB. Over time, however, users requested an independent version of AGVis to be made since it seemed useful outside of just being ANDES's visualizer. Creating a version of AGVis that does not rely on DiME and ANDES would align better with the LTB's focus of offering software that worked both as a cohesive set in a digital twin environment and independently. A large update was made to AGVis that added features to work independently from other applications on top of the original feature set. To put it simply, depending on how a user started AGVis, they would have two different feature sets to work with. Due to the lack of parity between the independent and dependent features, we will refer to them as two different entities throughout this section for the sake of clarity. Terms like "original" and "base" will refer to the AGVis features that require other programs. "MultiLayer" will refer to the features that are available with just AGVis's files.

*A. Base geovisualization*

Base AGVis requires some setup to use. The most common way to start to use it is by downloading and installing Docker [19], downloading the files for ANDES, DiME, and AGVis, and creating a Docker image to run and host the three applications simultaneously. Depending on a user's environment, they might have to perform additional setup. Once a user opens a browser and connects to wherever AGVis is being hosted, they will see a map. AGVis relies on Leaflet [20], a JavaScript library for creating interactive maps, to display power systems. The original version of AGVis uses five primary layers: the Tile layer, the Zone layer, the Topology layer, the Contour layer, and the Search layer. The data for the layers and the layers themselves are stored within a Window class. The Tile layer is the lowest layer and displays the world map, the Zone layer displays predefined regions such as WECC, EI or ERCOT, the Topology layer displays nodes and lines from simulation data on the map, the Contour layer animates a heat map from simulation data, and the Search layer allows nodes to be searched. Of these, the most important are the Topology and Contour layers.

When base AGVis connects to a DiME server, it waits until it begins to receive data. The data it receives is almost exclusively used by the Topology and Contour layers since they are employed to display the simulation animation. One of the first things AGVis receives is the topology data. Once it receives the data, AGVis parses it and displays the buses and lines of a given power system on the Topology layer. After that it will display a heat map of the simulation data as fast as AGVis receives it. This heat map is shown on the Contour layer. It is created by performing Delaunay triangulation on the points in the Topology layer and calculating the heat map for the triangles based on the simulation data as input to the Topology layer. Then it finally performs a smoothing function to create a gradient across the triangles. The Contour layer is updated every frame that new data is received.

Once a simulation stops sending data to AGVis, it will add a playback bar for the simulation. The playback bar uses the "history" object, which stores all data that was passed to AGVis. Users can view the simulation at its real time speed or at customizable speed. Users can also change certain parameters for the simulation, like which variable the heat map uses, how sensitive the heat map is to change, or where the timer starts and how much increments per second. If a user wants to change which simulation they are viewing, they must restart AGVis and ANDES.

*B. Multilayer feature*

The MultiLayer implementation of AGVis helps address a few of the problems with the original AGVis. Base AGVis requires users to restart two pieces of software to examine other simulations. Furthermore, its installation changes depending on a user's operating system. MultiLayer AGVis circumvents both issues by allowing users to upload more than one data set at a time and by only needing itself, a browser, and a way to locally host files to run. Although the MultiLayer features are available when running base AGVis, they are also available if a user only locally hosts AGVis's files. This means that the MultiLayer features can be run in effectively the same way regardless of operating system.

The MultiLayer functionality in AGVis uses a new type of Topology layer alongside the original. Instead of receiving data from DiME like the original Topology layer, this new Topology layer, called "MultiTop", uses data from a user-provided XLSX file to display nodes and lines. AGVis expects the XLSX file to be formatted in the same way as an XLSX file intended for ANDES. For each file a user uploads in the "Add Layers" menu of AGVis, an object called "newlayer" is created. The newlayer is MultiLayer's equivalent to the Window class of base AGVis. Each newlayer stores the file data and MultiTop associated with a given system. When a MultiTop layer is instantiated and added to the map, its rendering function uses specific variables from the newlayer to plot the system's nodes and lines. This approach of giving each new file its own layer and data object allows the MultiLayer implementation to go far beyond the original AGVis in customization and ease of use.

Giving each data set its own layer allows for users to customize a specific data set without affecting all the other data sets. MultiLayer is able to customize node colors, opacities, and sizes, and line colors, opacities, and sizes for each uploaded system. It can also freely add, hide, and delete MultiTop layers. The primary limitation of the MultiLayer implementation, and the main reason why base AGVis is still relevant, is that MultiLayer has no Contour layer equivalent for MultiTop layers. MultiLayer AGVis is currently unable to produce animations from power system simulations.

IV. CASE STUDIES

*A. Federal use of DiME, AGVis, and ANDES*

Although DiME is primarily used as a messaging environment for distributed computing, it also has uses as a simple way to pass data between programs of different languages. This is best shown when running base AGVis with ANDES. When they start, AGVis and ANDES both connect to a DiME server if one is available. Then, as shown in Fig. 3, DiME acts as an intermediary between them. ANDES interprets an XLSX file, PSS/E RAW file, or a DYR file and then sends data to the group that AGVis has joined. AGVis, meanwhile, waits until data is sent to it. AGVis then animates the simulation based on the data it receives. Fig. 4 shows AGVis displaying the topology for the CURENT North America test case. Fig. 5 shows screenshots from AGVis while in a dynamic simulation study, which shows the frequency contour map during the transient stability study. Fig. 5 contains a capture from both during and after the AGVis receives simulation data from ANDES.

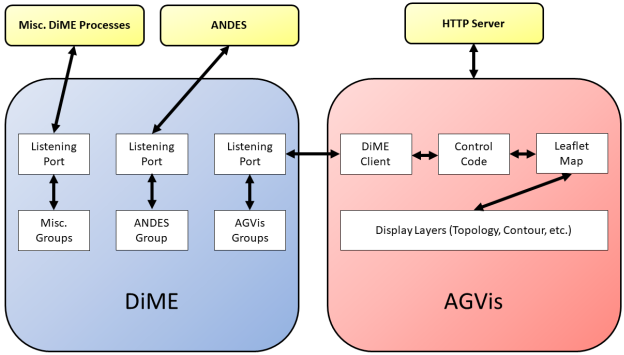

Figure 3. A component diagram of the interactions between ANDES, DiME, and AGVis.

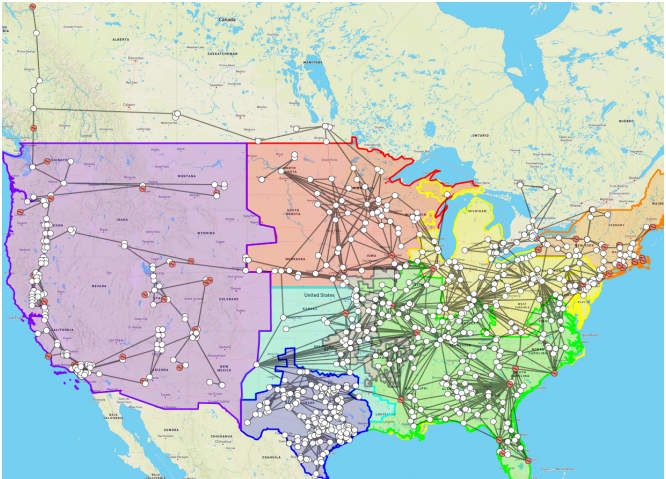

Figure 4. AGVis visualizing the entire NA system. UI cropped for clarity.

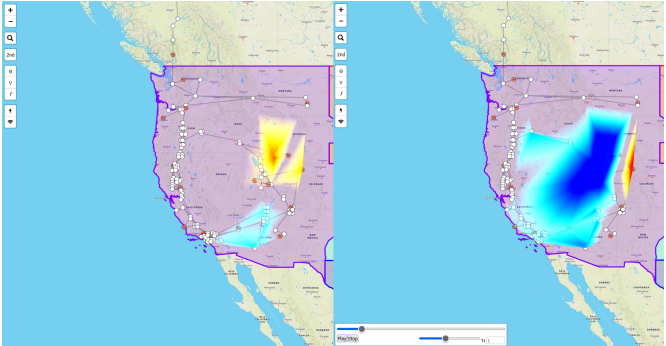

Figure 5. AGVis visualizing the WECC system. On the left is AGVis while receiving data from ANDES. On the right is AGVis after finishing data reception. Note the playback bar in the bottom left of the right image.

These animations are performed in real-time as AGVis receives data from DiME. Furthermore, the playback bar for reviewing a given animation becomes available as soon as DiME finishes sending data. This means users can interpret simulations at the speed that they run. The federal use of ANDES, DiME, and AGVis acts as a digital twin and provides a fast and distinct method for comparing differences when developing simulations for a given system.

### B. MultiLayer Implementation

The MultiLayer implementation of AGVis is best used for its customization options and convenience when adding new data, such as new buses. Fig. 6 shows AGVis after uploading two systems, an IEEE 39 bus system, and an example gas network. AGVis displays the transmission lines and buses for both systems. Both systems' nodes and lines have been customized. The "Prioritize Layer" button has been used on the IEEE 39 bus system so that it is rendered above the gas network. The other figure, Fig. 7, shows AGVis after deleting the IEEE system, leaving the example gas network. Fig. 6 and Fig. 7, both include the "Add Layers" menu to show how the menu accommodates the adding and removing of data.

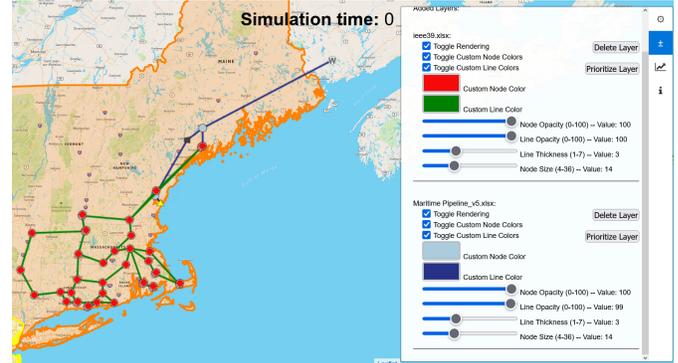

Figure 6. AGVis displaying the IEEE 39 bus and example gas network systems with customized settings.

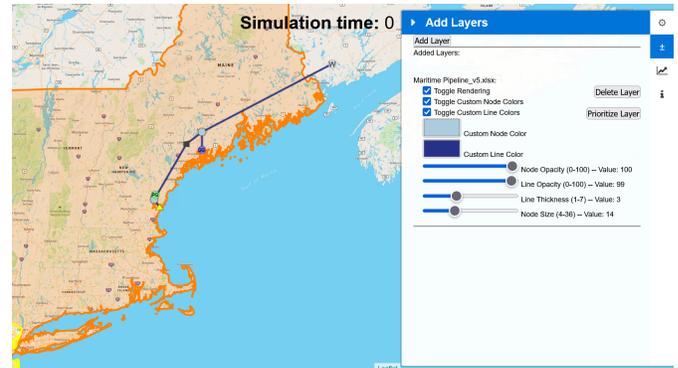

Figure 7. AGVis displaying only the gas network after deleting the IEEE 39 bus system.

The MultiLayer features of AGVis allow for users to efficiently create maps with distinct multi-energy systems using MultiLayer's customization options. Users can also use MultiLayer in tandem with the federal use of AGVis to examine where a simulated system might affect other, overlapping systems.

Additional demonstrations of both the MultiLayer and federal use of AGVis can be found on the official CURENT LTB YouTube channel [21]-[22].

### V. CONCLUSIONS

Large-scale power system simulation faces challenges involving high-concurrent data exchange and results interpretation. This paper presents two important packages in CURENT LTB: DiME, a distributed messaging environment and AGVis, a geovisualization tool. DiME implements a shared workspace model for distributed computing that can handle high-concurrency, high-volume data for JavaScript, Python, and MATLAB programs. With modular structure, the

LTB allows convenient, federal use of ANDES, DiME, and AGVis for real-time large-scale power system simulation and visualization as a digital twin. Further, with the MultiLayer feature, AGVis also facilitates user-provided data mapping. The results show that with the modular structure, LTB is capable of not only federal use for real-time large-scale power system simulation, but also independent use for various customized power system research.

In the future, DiME and AGVis will be further improved to be compatible with multi-energy systems and other emerging power system technologies. Development for AGVis in particular will focus on giving it a MultiLayer equivalent for the base version's ContourLayer so that it can display heatmap animations. Further testing will also be done with both programs to ensure that they are compatible and maintain the same ease of use with power systems beyond those in North America.

ACKNOWLEDGMENT

The authors would like to acknowledge the financial support from CURENT, a National Science Foundation (NSF) Engineering Research Center funded by NSF and Department of Energy under NSF Award EEC-1041877.